\documentclass[12pt]{article}
\usepackage[numbers]{natbib}
\usepackage{color}
\usepackage{times}
\usepackage{geometry}
\usepackage[utf8]{inputenc}
\usepackage{enumitem,amssymb}
\usepackage{ragged2e}
\newlist{thematic}{itemize}{8}
\setlist[thematic]{label=$\square$}
\usepackage{pifont}
\usepackage{wasysym}
\usepackage{graphicx}
\usepackage[font=footnotesize,labelfont=bf]{caption}
\usepackage{subcaption}
\usepackage{url}
\usepackage{wrapfig}

\begin{document}
{\raggedright
\Large
Astro2020 Science White Paper \linebreak

\huge
Radio Counterparts of Compact Object Mergers in the Era of Gravitational-Wave Astronomy \linebreak

\normalsize
\noindent \textbf{Thematic Areas:} \hspace*{60pt} $\square$ Planetary Systems \hspace*{10pt} $\square$ Star and Planet Formation \hspace*{20pt}\linebreak
$\XBox$ Formation and Evolution of Compact Objects \hspace*{31pt} $\square$ Cosmology and Fundamental Physics \linebreak
  $\square$  Stars and Stellar Evolution \hspace*{1pt} $\square$ Resolved Stellar Populations and their Environments \hspace*{40pt} \linebreak
  $\square$    Galaxy Evolution   \hspace*{45pt} $\XBox$             Multi-Messenger Astronomy and Astrophysics \hspace*{65pt} \linebreak
  
\textbf{Principal Author:}

Name:	Alessandra Corsi
 \linebreak						
Institution:  Texas Tech University
 \linebreak
Email: alessandra.corsi@ttu.edu
 \linebreak
Phone:  806-834-6931
 \linebreak
 
\textbf{Co-authors:} \\Nicole~M.~Lloyd-Ronning (Los Alamos National Lab, University of New Mexico, Los Alamos), Dario Carbone (Texas Tech University), Dale~A.~Frail (National Radio Astronomy Observatory), Davide Lazzati (Oregon State University), Eric~J.~Murphy (University of Virginia, National Radio Astronomy Observatory), Richard O'Shaughnessy (Rochester Institute of Technology), Benjamin~J.~Owen (Texas Tech University), David~J.~Sand (University of Arizona), Wen-Fai Fong (Northwestern University), Kristine Spekkens (Royal Military College of Canada), Andrew Seymour (Associated Universities, Inc.).
 \linebreak

\textbf{Abstract:}
GHz radio astronomy has played a fundamental role in the recent dazzling discovery of GW170817, a neutron star (NS)-NS merger observed in both  gravitational waves (GWs) and light at all wavelengths. Here we show how the expected progress in sensitivity of ground-based GW detectors over the next decade calls for U.S.-based GHz radio arrays to be improved beyond current levels.  We discuss specifically how several new scientific opportunities would emerge in multi-messenger time-domain astrophysics if a next generation GHz radio facility with sensitivity and resolution $10\times$ better than the current Jansky Very Large Array (VLA) were to work in tandem with ground-based GW detectors. These opportunities include probing the properties, structure, and size of relativistic jets and wide-angle ejecta from NS-NS mergers, as well as unraveling the physics of their progenitors via host galaxy studies. }

\thispagestyle{empty}

\pagebreak

\pagenumbering{arabic}

{\Large{\bf Introduction}}

 With the discovery of gravitational waves (GWs) from compact object mergers, a new window to the universe has been opened, providing an opportunity to make enormous gains in our understanding of some of the most exotic objects in the stellar graveyard -- black holes (BHs) and neutron stars (NSs). During their first two observing runs (O1/O2), the advanced Laser Interferometer Gravitational wave Observatory (LIGO) and Virgo detected several BH-BH binaries (see e.g., \citep{LIGO2018Catalog} and references therein),  as well as GW170817, a NS-NS merger accompanied by a $\gamma$-ray burst (GRB) plus an afterglow spanning all bands of the electromagnetic (EM) spectrum (see e.g., \citep{GW170817Discovery,GW170817GRBDiscovery,GW170817EMFollow,Coulter2017,Evans2017,Hallinan2017,Kasliwal2017,Troja2017,Valenti2017}).
 The broad-band EM follow up of GW170817 was key to providing arsec localization and to probing different facets of the merger. While optical/IR observations revealed a kilonova and the merger as a site of r-process nucleosynthesis (see \citep{Barnes2013,Metzger2017} and references therein), radio and X-rays probed a completely different component, namely, a fast ejecta with slower wings observed at large angles  (e.g., \citep{Mooley2017,Alexander2018,Lazzati2018,Dobie2018,Mooley2018,MooleyVLBA}). The Karl G. Jansky Very Large Array (VLA; \citep{VLAAcknow}), in particular, played a fundamental role in unveiling the relatively faint GHz counterpart of GW170817 \citep{Hallinan2017}, and subsequently providing definitive evidence for the presence of a fast, structured jet launched after the merger \citep{Lazzati2018,Mooley2018,Lazzati2017}.
 
 Here we discuss expectations for how GHz radio observations will continue to contribute to the new field of GW astronomy over the next decade. Ten years from now the network of ground-based GW detectors is projected to include Virgo operating at nominal advanced sensitivity, the two advanced LIGO detectors likely in their so-called ``plus configuration'' (which foresees up to a factor of $\sim7$ increase in volume of space surveyed; \citep{NSFPress}), the Kamioka Gravitational Wave Detector (KAGRA), and LIGO India \citep{LIGOLivRev}. This network of detectors will be identifying a large number of GW in-spirals and mergers in the local universe \citep{NSFPress,LIGOLivRev}.  
 We make the case that the projected increase in sensitivity of GW detectors over the next decade dramatically calls for an improvement in both sensitivity and spatial resolution on current U.S.-based GHz radio arrays. 
We support our case by discussing several avenues along which transformational results could be enabled if a facility with sensitivity and spatial resolution $\approx 10\times$ better than the current VLA were to work in tandem with ground-based GW detectors:
\begin{enumerate}	
\item{{\bf Physics of NS-NS/NS-BH jets via radio continuum observations}:   The details of the physics  behind relativistic jets produced in compact binary mergers -- in particular,  particle acceleration and dissipation mechanisms -- are not well understood and have implications for a wide variety of astrophysical outflows including Active Galactic Nuclei (AGN) and X-ray Burst (XRB) jets. Rapid radio follow up of NS-NS (and NS-BH) mergers in the local universe will help elucidate the physics of these outflows.}

\item{{\bf Ejecta structure and magnetic fields via VLBI and radio polarimetry}: Before the discovery of GW170817, we have studied relativistic jets produced in compact object mergers by relying largely on their highly beamed $\gamma$-ray emission (a.k.a. GRB) for discovery. Over the next decade, GW observatories will enable us to discover numerous NS-NS/NS-BH systems with off-axis jets (thus lacking a luminous $\gamma$-ray counterpart). By combining direct mapping (via VLBI) and radio polarimetry of off-axis jets, we can directly probe the distribution of ejecta speeds and the as-yet poorly understood configuration of their magnetic fields. These, in turn, are linked to open fundamental physics questions, such as the equation of state and magnetization of the merging compact objects.}

\item{{\bf Host galaxies of NS-NS/NS-BH mergers}: With radio observations of the host galaxies of NS-NS/NS-BH mergers, we can map the galaxy star formation rates (SFRs), constrain the fraction of late- versus early-type galaxies and the  star formation histories, measure the distribution of offsets of merger sites with respect to the host galaxy light, and ultimately help unravel formation scenarios for the progenitors of NS-NS/NS-BH systems.
}
\end{enumerate}

In what follows we discuss the above scientific opportunities in more detail, assuming the astronomical community will have access to a next generation radio telescope, hereafter referred to as ``next generation Very Large Array'' (ngVLA), characterized by a nominal 10-fold improvement in both sensitivity and resolution over the current VLA \citep{Corsi2018proc,Murphy2018}.  We note that a nominal next generation array such as the one discussed here would surpass the capabilities of facilities such as SKA1-MID \citep{SKA,Bourke2018} in terms of angular resolution and sensitivity at cm and shorter wavelengths, and would complement strongly SKA1 and ALMA by providing enhanced sensitivity and resolution to pursue new scientific goals.

 \begin{figure}
\begin{center}
\hbox{
\hspace{-0.6cm}
\includegraphics[height=7.1cm]{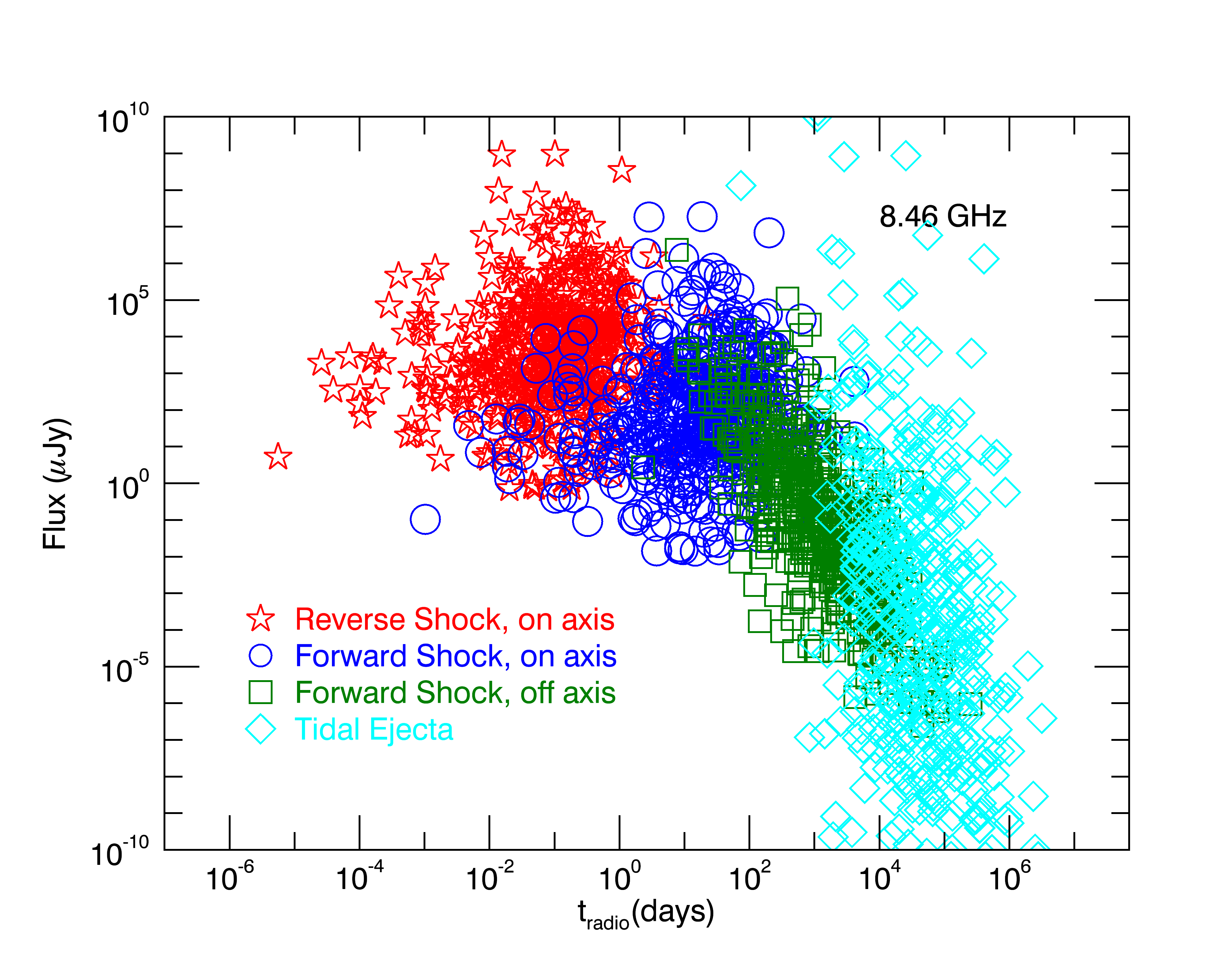}
\hspace{-0.8cm}
\includegraphics[height=6.8cm]{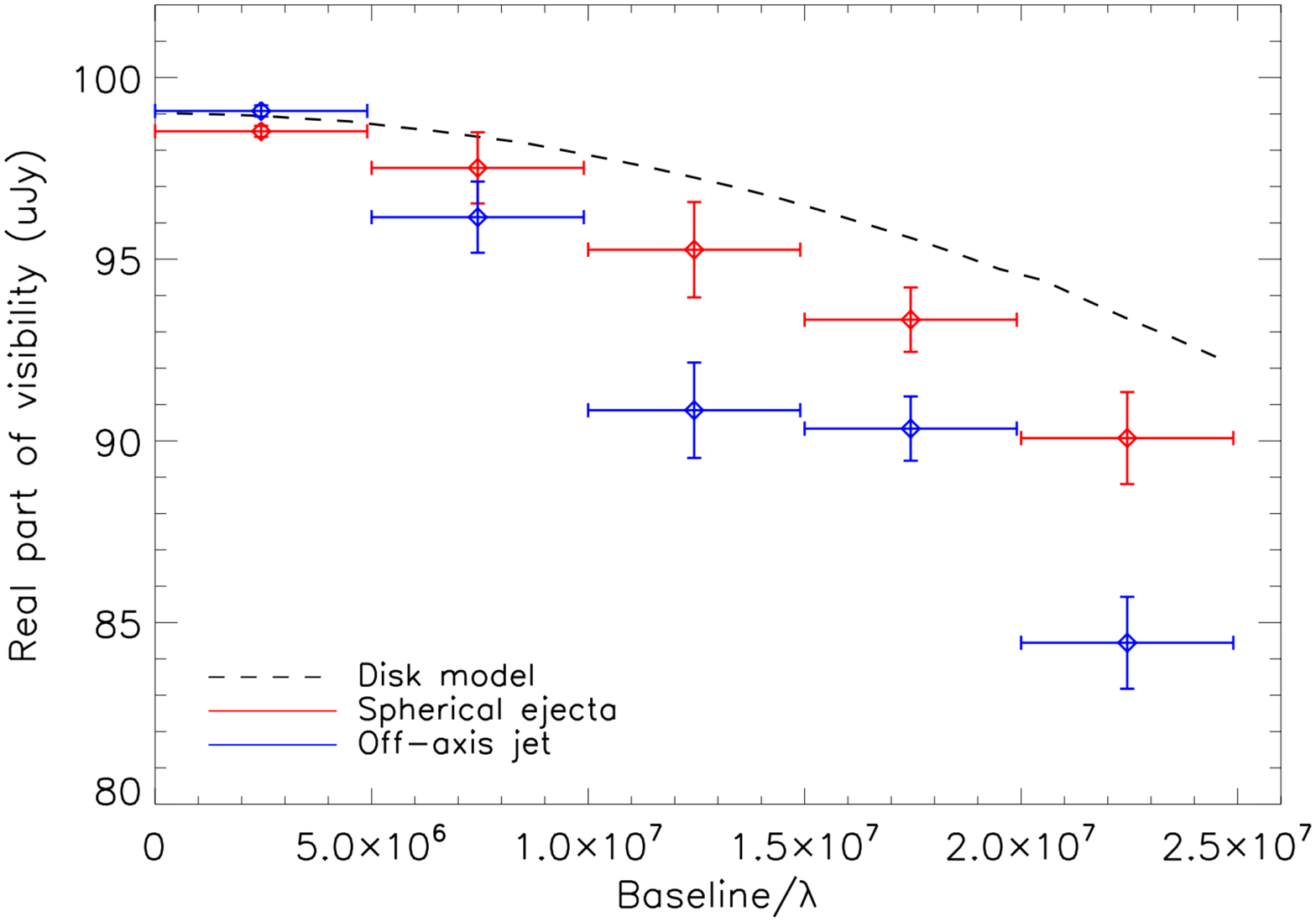}}
\end{center}
\vspace{-1.6cm}
\caption{\it LEFT: Radio flux density as a function of time for an on-axis forward shock (blue), reverse shock (red), off-axis forward shock (green), and tidal ejecta component (cyan) for a short GRB from a NS-NS merger.  We assume a mean $n_{ISM}=0.005$ cm$^{-3}$ with a large spread (2.5 orders of magnitude), a mean energy of $E = 10^{51}erg$, a mean redshift of $0.5$. $\epsilon_{e}$ and $\epsilon_{B}$ are both drawn from a Gaussian with a mean of 0.1. For reference, an ngVLA would reach a 3$\sigma$ sensitivity of $\approx 0.66\,\mu$Jy in 1\,hr at 8\,GHz. RIGHT: Real part of the ngVLA visibility as a function of baseline for two models (isotropic and off-axis collimated fireballs at 150\,days since explosion when the 2.4\,GHz flux density is of order $\approx 100\,\mu$Jy, comparable to the 3\,GHz peak flux density of GW170817) compared to that of a uniform disk of  $2\times10^{-3}$\,arcsec diameter and total flux density of 100$\,\mu$Jy. We assume a 4\,hr-long observation at a central frequency of 2.4\,GHz with an ngVLA-like radio array which could reach a noise rms of $\approx 0.2\,\mu$Jy at this frequency \citep{Corsi2018proc}. An ngVLA could directly resolve and distinguish different ejecta structures, something currently inaccessible to the VLA.}
\label{fig:Radfluxall}
\end{figure}

\smallskip\smallskip\smallskip
{\Large{\bf Physics of NS-NS/NS-BH jets}}

The long-term evolution of the radio afterglow of GW170817 has ultimately confirmed that short GRB-like jets can emerge successfully from NS-NS mergers, even though when misaligned with our line of sight (off-axis) they lack a bright, fast-decaying afterglow \citep{Lazzati2018,Mooley2018,MooleyVLBA}.  Overall, with only one off-axis NS-NS merger probed so far, the question of whether successful relativistic jets are formed in \textit{all} binary NS mergers remains open. Answering this question is crucial to understanding whether short GRBs track the NS-NS merger rate, or if instead a larger variety of ejecta outcomes is possible in these mergers, including so-called  ``choked'' jets (see e.g. \citep{Hallinan2017,Kasliwal2017,Nakar2018} for a detailed discussion).  Building a large sample of NS-NS/NS-BH mergers with well sampled radio counterparts is key to this end, but can only be accomplished with enough sensitivity in the GHz radio band (Figure 1, left). Indeed, the radio afterglow of GW170817, with a GHz peak flux density of order $\approx 100\,\mu$Jy, would quickly become undetectable with the current VLA if placed at the NS-NS distance horizon expected for near future runs of even just \textit{current} GW detectors ($\sim 120$\,Mpc; \citep{LIGOLivRev}). 

In the standard picture of a relativistic external blast wave \citep{Meszaros2006}, the onset of the GRB afterglow occurs around the deceleration time -- i.e. when the blast wave has swept up enough external material to decelerate. At that time, a reverse shock also crosses the shocked shell material giving rise to short-lived emission.
The characteristic synchrotron frequencies of the afterglow spectrum ($\nu_{a}$, the self-absorption frequency, $\nu_{m}$, the frequency corresponding to the ``minimum'' characteristic electron energy, and $\nu_{c}$, the frequency at which an electron loses most of its energy to radiation), can constrain key explosion and environment parameters such as the total kinetic energy ($E$), the fraction of energy in the electric and magnetic fields ($\epsilon_{e}$ and $\epsilon_{B}$), and the density of the external medium ($n_{ISM}$).

The radio band is crucial for pinning down these frequencies and understanding the physics behind relativistic outflows.  
As shown in Figure 1 (left), a facility like the ngVLA would give us the opportunity for the first time to catch the peak of the reverse shock emission (which remained unprobed in GW170817), breaking the degeneracies among the various physical parameters that determine GRB emission from NS-NS (and NS-BH) mergers. Specifically, radio observations from an ngVLA  would distinguish between reverse and forward shock emission based on the time at which the radio band emission peaks. More detailed modeling of the time evolution of the radio emission would provide further detailed constraints on the jet physics and ISM properties, probing GW170817-like emission up to distances $\approx 3\times$ larger than currently accessible with the VLA (which implies a $30\times$ larger event rate; see \citep{Carbone2017} for a more detailed discussion). 
  

\smallskip\smallskip\smallskip
{\Large{\bf NS-NS/NS-BH ejecta and magnetic field structure}}

\begin{figure}
\begin{center}
\hbox{
\hspace{-0.5cm}
\includegraphics[height=6.8cm]{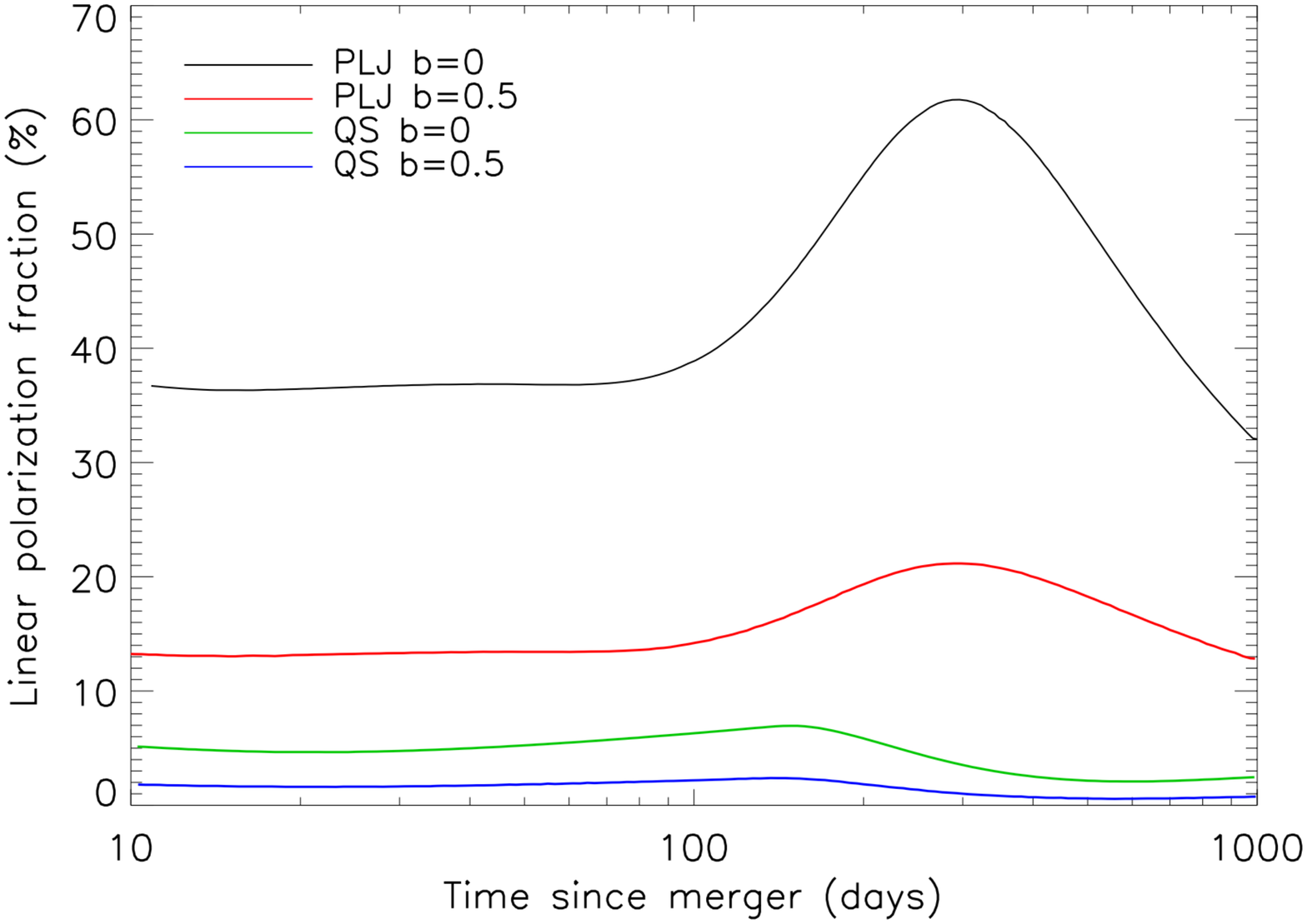}
\hspace{-0.6cm}
\includegraphics[height=6.8cm]{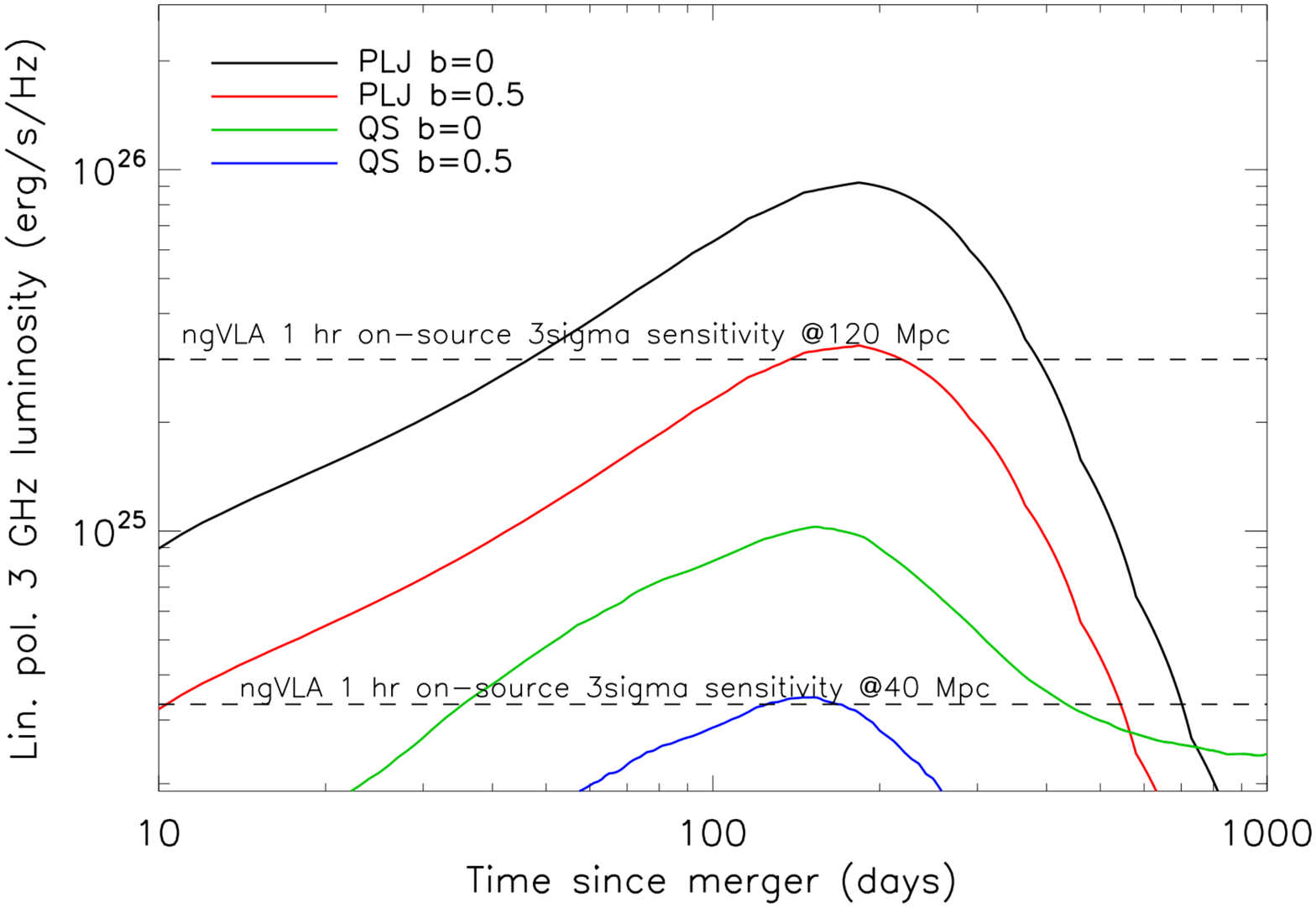}}
\vspace{-0.5cm}
\caption{\it LEFT: Linear polarization fractions for different ejecta structures (power-law jet, PLJ, versus quasi-spherical, QS) and magnetic field configurations (fully contained in the plane of the shock for $b=0$) as predicted by \citep{Granot2018}. RIGHT: Expected amount of linearly polarized luminosity density at 3\,GHz for a GW170817-like afterglow given the different degrees of linear polarization shown in the left panel. A radio array with the sensitivity of the ngVLA could probe above the dashed horizontal lines for mergers located at 40/120\,Mpc. Thus, an ngVLA would be providing us with a detailed view of a large variety of ejecta and magnetic field structures, that are inaccessible to current radio facilities \citep{Corsi2018proc,Corsi2018}.  \label{radiolight}}
\end{center}
\vspace{-1.cm}
\end{figure}
Although continuum radio emission properties can constrain the angle under which we happen to be observing NS-NS/NS-BH jets (see previous Section), resolved imaging of the radio ejecta associated with NS-NS/NS-BH mergers using VLBI techniques represents the only direct way to map the speed distribution of merger ejecta, and distinguish unambiguously e.g. collimated relativistic fireballs observed off-axis from quasi-spherical relativistic ejecta \citep{MooleyVLBA,Nakar2018}. 
The radio image of NS-NS merger ejecta will generally depend on the details of the interaction of the fastest ejecta component with the slower, neutron-rich material, and several different outcomes are possible \citep{Nakar2018}. 
As demonstrated in the case of GW170817, off-axis collimated ejecta will appear to have a time-dependent emission centroid increasingly offset from the location of the counterpart that one would measure at early times \citep{MooleyVLBA}. With an ngVLA with $\sim 30\times$ longer baselines ($\sim 1000$\,km) than the current VLA (plus potentially extended baselines to continental scales of $\approx 8860$\,km) one would not only be able to build an accurate map of the time evolution of such an offset over time, but also and more importantly resolve directly radio ejecta of NS-NS (and NS-BH) mergers as dim as GW170817 ($\approx 100\,\mu$Jy). This is demonstrated in the right panel of Figure 1, where we show how an analysis of the radio visibility measured by an ngVLA as a function of baseline will distinguish off-axis collimated outflows from isotropic ones \citep{Corsi2018proc}.

 Direct radio imaging, paired with radio polarimetry, can also provide a strong discriminant of ejecta geometry and magnetic field structure \citep{Granot2018,Corsi2018,Rossi2004}.  
This is demonstrated in the left panel of Figure 2, where we show how different magnetic field configurations within a NS-NS/NS-BH ejecta give rise to different degrees of linearly polarized radio emission, as predicted by \citep{Granot2018}. Specifically, if the magnetic field is completely tangled in the plane of the shocked ejecta, a large degree of linear polarization should be observed and could be used as a smoking gun for the presence of a successful misaligned jet accompanying a large-angle outflow in a NS-NS merger.  On the other hand, a quasi-spherical outflow that could result from a choked jet would produce linearly polarized radio emission at a much lower level. 
As evident from the right panel of Figure 2, we need the sensitivity of an ngVLA to probe the time-dependent degree of linear polarization of GW170817-like radio counterparts up to $d_L\sim 120$\,Mpc -- the near future horizon distance for NS-NS mergers of ground-based GW detectors \citep{LIGOLivRev}.

\smallskip\smallskip\smallskip
{\Large{\bf Host Galaxies}}

Understanding the progenitors and formation channels of compact binary systems (i.e. how they actually form and evolve) requires the identification of host galaxies, and a detailed study of the merger environment. The distribution of merger timescales impacts the mix of early- and late-type hosts \citep{Richard2008,Richard2010}, and the level of correlation with (recent) star formation rather than with stellar mass alone (e.g., \citep{Zheng2007}). With radio studies of both local NS-NS/NS-BH mergers (detected via GWs)  and short GRBs at higher redshifts, we can build a consistent picture of the local and galactic environments in which mergers occur, taking into account the evolving galaxy population.

Radio continuum emission from galaxies traces the SFR during the last $\sim 50-100$\,Myr (e.g., \citep{Murphy2011}), and could thus help track the connection between mergers and recent SFR. In fact, due to the age-dust-metallicity degeneracy, the optical/UV emission alone is an unreliable measure of SFR in dusty galaxies. Of course, radio measurements are subject to AGN contamination.  That said, radio measurements have been demonstrated to be an important complementary tool to constrain the SFR in the past, for example for short GRB hosts (e.g., \citep{Nicuesa2014}).  
\begin{wrapfigure}{r}{0.55\textwidth}
\vspace{-30pt}
\begin{center}
\hspace{-30pt}
   \includegraphics[height=7.cm]{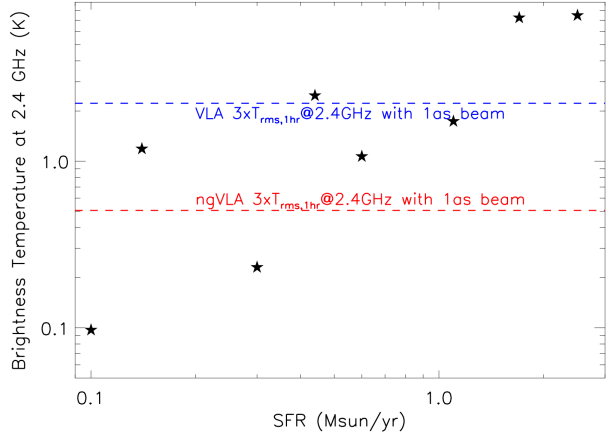}
   \end{center}
   \vspace{-37pt}
   \caption{\it Radio brightness temperature of NS-NS/NS-BH merger hosts at 120\,Mpc \citep{Corsi2018proc}, extrapolated from the current sample of short GRB hosts (e.g., \citep{Berger2014}). An ngVLA will resolve most of these hosts (all of the ones located above the red dashed line), while more than half would be inaccessible to the current VLA (the latter would only be able to probe above the blue line).}
   \vspace{-21pt}
\end{wrapfigure}
Now with local NS-NS mergers discovered via GWs we can trace the immediate gas/environment also using single radio dishes such as the Green Bank Telescope (GBT). Indeed for NGC4993, the host galaxy of GW170817, a critical environmental measurement  came  from  a  non-detection of HI gas by the GBT \citep{Hallinan2017}, which ruled  out  potentially  obscured  star  formation,  lending  credence to  the  optical/NIR  spectral energy distribution  and  host  spectral  modeling  efforts  \citep{Blanchard2017,Levan2017}. Studies of atomic gas with single-dish radio telescopes are thus complementary to investigations with interferometric arrays. Moreover, when phased with large single-dish telescopes, the sensitivity and resolution of radio arrays can be further improved. An ngVLA will be crucial to carry out resolved studies of compact binary merger hosts \citep{Fong2013}, probing the presence or absence of a spatial association with star formation or stellar mass, and constraining the distribution of offsets of electromagnetic counterparts with respect to the host galaxy light (which maps natal kicks, e.g. \citep{Belczynski2006,Behroozi2014}). 
Indeed, in Figure 3 we demonstrate how an ngVLA will resolve short GRB-like hosts within $\approx 120$\,Mpc. 

We finally note that statistical associations between a population of GW events and potential host galaxies can also be leveraged for compact object mergers without confident electromagnetic counterparts (thus lacking an arcsec localization).  In this respect, radio observations of host galaxies are likely to become relevant also in population studies of BH-BH mergers \citep{Richard2017}. 

\smallskip\smallskip\smallskip
{\Large{\bf Summary}}

 If we are to truly exploit the multi-messenger nature of the universe's most energetic and exotic events, radio observations with an ngVLA ($10$-fold improvement in sensitivity and angular resolution compared to the current VLA) are necessary. \textbf{Without a future facility like the ngVLA, current U.S.-based radio interferometric arrays would quickly become uncompetitive for enabling the radio follow-up of GW170817-like NS-NS mergers detected by increasingly sensitive ground-based GW detectors.} Because GHz radio counterparts of NS-NS/NS-BH mergers uniquely probe the fastest merger outflow components, crucial information would be lost without radio observations. In contrast to X-ray wavelengths (which can also probe fast ejecta), radio emission is not strongly suppressed by viewing angle and relativistic beaming effects, and as such it represents a critical window of the electromagnetic spectrum for allowing us to map the diversity in ejecta structure of NS-NS/NS-BH mergers. Radio polarization can probe the structure of magnetic fields, something largely inaccessible to current interferometric radio arrays and key for constraining the equation of state and magnetization of the merging compact objects. Finally, as the sensitivity of ground-based GW detectors improves and more events are detected, an ngVLA  would enable population studies of compact object mergers and their hosts, providing invaluable clues on how compact binary systems form and evolve, how often they merge, and what is the zoo of possible merger outcomes in relation to the properties of the merger environment. 

 

\pagebreak

\bibliographystyle{unsrt}
\bibliography{decadal.bib}

\begin{thebibliography}{10}

\bibitem{LIGO2018Catalog}
B.~P. {Abbott} et~al.
\newblock {GWTC-1: A Gravitational-Wave Transient Catalog of Compact Binary
  Mergers Observed by LIGO and Virgo during the First and Second Observing
  Runs}.
\newblock {\em arXiv e-prints}, page arXiv:1811.12907, 2018.

\bibitem{GW170817Discovery}
B.~P. {Abbott} et~al.
\newblock {GW170817: Observation of Gravitational Waves from a Binary Neutron
  Star Inspiral}.
\newblock {\em Physical Review Letters}, 119(16):161101, 2017.

\bibitem{GW170817GRBDiscovery}
B.~P. Abbott et~al.
\newblock Gravitational waves and gamma-rays from a binary neutron star merger:
  {GW}170817 and {GRB} 170817a.
\newblock {\em The Astrophysical Journal}, 848(2):L13, 2017.

\bibitem{GW170817EMFollow}
B.~P. {Abbott} et~al.
\newblock {Multi-messenger Observations of a Binary Neutron Star Merger}.
\newblock {\em The Astrophysical Journal}, 848:L12, 2017.

\bibitem{Coulter2017}
D.~A. {Coulter} et~al.
\newblock {Swope Supernova Survey 2017a (SSS17a), the optical counterpart to a
  gravitational wave source}.
\newblock {\em Science}, 358:1556--1558, 2017.

\bibitem{Evans2017}
P.~A. {Evans} et~al.
\newblock {Swift and NuSTAR observations of GW170817: Detection of a blue
  kilonova}.
\newblock {\em Science}, 358:1565--1570, 2017.

\bibitem{Hallinan2017}
G.~{Hallinan}, A.~{Corsi}, et~al.
\newblock {A radio counterpart to a neutron star merger}.
\newblock {\em Science}, 358:1579--1583, 2017.

\bibitem{Kasliwal2017}
M.~M. {Kasliwal} et~al.
\newblock {Illuminating gravitational waves: A concordant picture of photons
  from a neutron star merger}.
\newblock {\em Science}, 358:1559--1565, 2017.

\bibitem{Troja2017}
E.~{Troja} et~al.
\newblock {The X-ray counterpart to the gravitational-wave event GW170817}.
\newblock {\em Nature}, 551:71--74, 2017.

\bibitem{Valenti2017}
S.~{Valenti} et~al.
\newblock {The Discovery of the Electromagnetic Counterpart of GW170817:
  Kilonova AT 2017gfo/DLT17ck}.
\newblock {\em The Astrophysical Journal Letters}, 848:L24, 2017.

\bibitem{Barnes2013}
J.~{Barnes} and D.~{Kasen}.
\newblock {Effect of a High Opacity on the Light Curves of Radioactively
  Powered Transients from Compact Object Mergers}.
\newblock {\em The Astrophysical Journal}, 775:18, 2013.

\bibitem{Metzger2017}
B.~D. {Metzger}.
\newblock {Welcome to the Multi-Messenger Era! Lessons from a Neutron Star
  Merger and the Landscape Ahead}.
\newblock {\em arXiv e-prints}, page arXiv:1710.05931, 2017.

\bibitem{Mooley2017}
K.~P. {Mooley} et~al.
\newblock {A mildly relativistic wide-angle outflow in the neutron-star merger
  event GW170817}.
\newblock {\em Nature}, 554:207--210, 2018.

\bibitem{Alexander2018}
K.~D. {Alexander} et~al.
\newblock {A Decline in the X-Ray through Radio Emission from GW170817
  Continues to Support an Off-axis Structured Jet}.
\newblock {\em The Astrophysical Journal Letters}, 863:L18, 2018.

\bibitem{Lazzati2018}
D.~{Lazzati} et~al.
\newblock {Late Time Afterglow Observations Reveal a Collimated Relativistic
  Jet in the Ejecta of the Binary Neutron Star Merger GW170817}.
\newblock {\em Physical Review Letters}, 120:241103, 2018.

\bibitem{Dobie2018}
D.~{Dobie} et~al.
\newblock {A Turnover in the Radio Light Curve of GW170817}.
\newblock {\em The Astrophysical Journal Letters}, 858:L15, 2018.

\bibitem{Mooley2018}
K.~P. {Mooley} et~al.
\newblock {A Strong Jet Signature in the Late-time Light Curve of GW170817}.
\newblock {\em The Astrophysical Journal Letters}, 868:L11, 2018.

\bibitem{MooleyVLBA}
K.~P. {Mooley} et~al.
\newblock {Superluminal motion of a relativistic jet in the neutron-star merger
  GW170817}.
\newblock {\em Nature}, 561:355--359, 2018.

\bibitem{VLAAcknow}
{The National Radio Astronomy Observatory is a facility of the National Science
  Foundation operated under cooperative agreement by Associated Universities,
  Inc.}

\bibitem{Lazzati2017}
D.~{Lazzati} et~al.
\newblock {Off-axis emission of short {$\gamma$}-ray bursts and the
  detectability of electromagnetic counterparts of gravitational-wave-detected
  binary mergers}.
\newblock {\em Monthly Notices of the Royal Astronomical Society},
  471:1652--1661, 2017.

\bibitem{NSFPress}
{National Science Foundation}.
\newblock Press statement: Upgraded {LIGO} to search for universe's most
  extreme events.
\newblock \url{https://www.nsf.gov/news/news_summ.jsp?cntn_id=297414}, 2019.

\bibitem{LIGOLivRev}
B.~P. Abbott et~al.
\newblock Prospects for observing and localizing gravitational-wave transients
  with advanced ligo and advanced virgo.
\newblock {\em Living Reviews in Relativity}, 19(1):1, 2016.

\bibitem{Corsi2018proc}
A.~{Corsi} et~al.
\newblock {Compact Binary Mergers as Traced by Gravitational Waves}.
\newblock In E.~{Murphy}, editor, {\em Science with a Next Generation Very
  Large Array}, volume 517 of {\em Astronomical Society of the Pacific
  Conference Series}, page 689, 2018.

\bibitem{Murphy2018}
E.~J. {Murphy} et~al.
\newblock {The ngVLA Science Case and Associated Science Requirements}.
\newblock In Eric {Murphy}, editor, {\em Science with a Next Generation Very
  Large Array}, volume 517 of {\em Astronomical Society of the Pacific
  Conference Series}, page~3, 2018.

\bibitem{SKA}
D.~R. {DeBoer} et~al.
\newblock {Australian SKA Pathfinder: A High-Dynamic Range Wide-Field of View
  Survey Telescope}.
\newblock {\em IEEE Proceedings}, 97:1507--1521, 2009.

\bibitem{Bourke2018}
T.~{Bourke} et~al.
\newblock {Expected Science Performance of the Square Kilometre Array Phase 1
  (SKA1)}.
\newblock In {\em American Astronomical Society Meeting Abstracts \#231},
  volume 231 of {\em American Astronomical Society Meeting Abstracts}, page
  152.07, 2018.

\bibitem{Nakar2018}
E.~{Nakar} et~al.
\newblock {From {\ensuremath{\gamma}} to Radio: The Electromagnetic Counterpart
  of GW170817}.
\newblock {\em The Astrophysical Journal}, 867:18, 2018.

\bibitem{Meszaros2006}
P.~{M{\'e}sz{\'a}ros}.
\newblock {Gamma-ray bursts}.
\newblock {\em Reports on Progress in Physics}, 69:2259--2321, 2006.

\bibitem{Carbone2017}
D.~{Carbone} and A.~{Corsi}.
\newblock {Optimized Radio Follow-up of Binary Neutron-star Mergers}.
\newblock {\em The Astrophysical Journal}, 867:135, 2018.

\bibitem{Granot2018}
R.~{Gill} and J.~{Granot}.
\newblock {Afterglow imaging and polarization of misaligned structured GRB jets
  and cocoons: breaking the degeneracy in GRB 170817A}.
\newblock {\em Monthly Notices of the Royal Astronomical Society},
  478:4128--4141, 2018.

\bibitem{Corsi2018}
A.~{Corsi} et~al.
\newblock {An Upper Limit on the Linear Polarization Fraction of the GW170817
  Radio Continuum}.
\newblock {\em The Astrophysical Journal Letters}, 861:L10, 2018.

\bibitem{Rossi2004}
E.~M. {Rossi} et~al.
\newblock {The polarization of afterglow emission reveals {$\gamma$}-ray bursts
  jet structure}.
\newblock {\em Monthly Notices of the Royal Astronomical Society}, 354:86--100,
  2004.

\bibitem{Richard2008}
R.~{O'Shaughnessy} et~al.
\newblock {Short Gamma-Ray Bursts and Binary Mergers in Spiral and Elliptical
  Galaxies: Redshift Distribution and Hosts}.
\newblock {\em The Astrophysical Journal}, 675:566--585, 2008.

\bibitem{Richard2010}
R.~{O'Shaughnessy} et~al.
\newblock {Binary Compact Object Coalescence Rates: The Role of Elliptical
  Galaxies}.
\newblock {\em The Astrophysical Journal}, 716:615--633, 2010.

\bibitem{Zheng2007}
Z.~{Zheng} and E.~{Ramirez-Ruiz}.
\newblock {Deducing the Lifetime of Short Gamma-Ray Burst Progenitors from Host
  Galaxy Demography}.
\newblock {\em The Astrophysical Journal}, 665:1220--1226, 2007.

\bibitem{Murphy2011}
E.~J. {Murphy} et~al.
\newblock {Calibrating Extinction-free Star Formation Rate Diagnostics with 33
  GHz Free-free Emission in NGC 6946}.
\newblock {\em The Astrophysical Journal}, 737:67, 2011.

\bibitem{Nicuesa2014}
A.~{Nicuesa Guelbenzu}, S.~{Klose}, et~al.
\newblock Another short-burst host galaxy with an optically obscured high star
  formation rate: The case of grb 071227.
\newblock {\em The Astrophysical Journal}, 789(1):45, 2014.

\bibitem{Berger2014}
E.~{Berger}.
\newblock {Short-Duration Gamma-Ray Bursts}.
\newblock {\em Annual Review of Astronomy \& Astrophysics}, 52:43--105, 2014.

\bibitem{Blanchard2017}
P.~K. {Blanchard} et~al.
\newblock {The Electromagnetic Counterpart of the Binary Neutron Star Merger
  LIGO/Virgo GW170817. VII. Properties of the Host Galaxy and Constraints on
  the Merger Timescale}.
\newblock {\em The Astrophysical Journal Letters}, 848:L22, 2017.

\bibitem{Levan2017}
A.J. {Levan} et~al.
\newblock The environment of the binary neutron star merger gw170817.
\newblock {\em The Astrophysical Journal Letters}, 848:28, 2017.

\bibitem{Fong2013}
W.~{Fong} and E.~{Berger}.
\newblock The locations of short gamma-ray bursts as evidence for compact
  object binary progenitors.
\newblock {\em The Astrophysical Journal}, 776(1):18, 2013.

\bibitem{Belczynski2006}
K.~{Belczynski} et~al.
\newblock {A Study of Compact Object Mergers as Short Gamma-Ray Burst
  Progenitors}.
\newblock {\em The Astrophysical Journal}, 648:1110--1116, 2006.

\bibitem{Behroozi2014}
P.~S. {Behroozi} et~al.
\newblock {Interpreting Short Gamma-Ray Burst Progenitor Kicks and Time Delays
  using the Host Galaxy-Dark Matter Halo Connection}.
\newblock {\em The Astrophysical Journal}, 792:123, 2014.

\bibitem{Richard2017}
R.~{O'Shaughnessy} et~al.
\newblock {The effects of host galaxy properties on merging compact binaries
  detectable by LIGO}.
\newblock {\em Monthly Notices of the Royal Astronomical Society},
  464:2831--2839, 2017.

\end{thebibliography}



\end{document}